\documentclass[]{gSCS2e}

\usepackage{subfigure}
\usepackage{natbib,color}
\theoremstyle{plain}
\newtheorem{theorem}{Theorem}[section]
\newtheorem{corollary}[theorem]{Corollary}

\theoremstyle{definition}
\newtheorem{definition}[theorem]{Definition}

\theoremstyle{remark}

\begin{document}



\title{On the strengths of the self-updating process clustering algorithm}\thanks{Supported by NSC 96-2118-M-001-007-}

\author{Shang-Ying Shiu$^{\rm a}$$^{\ast}$
\vspace{6pt} and Ting-Li Chen$^{\rm b}$ \thanks{$^\ast$Corresponding author. Email: tlchen@stat.sinica.edu.tw}\\\vspace{6pt}  $^{a}${\em{Department of Statistics, National Taipei University, New Taipei, Taiwan}};
$^{b}${\em{Institute of Statistical Science, Academia Sinica, Taipei, Taiwan}}\\\received{September 2014} }
\maketitle

\begin{abstract}
The Self-Updating Process (SUP) is a clustering algorithm that stands from the viewpoint of data points and simulates the process how data points move and perform self-clustering. It is an iterative process on the sample space and allows for both time-varying and time-invariant operators. By simulations and comparisons, this paper shows that SUP is particularly competitive in clustering (i) data with noise, (ii) data with a large number of clusters, and (iii) unbalanced data. When noise is present in the data, SUP is able to isolate the noise data points while performing clustering simultaneously. The property of the local updating enables SUP to handle data with a large number of clusters and data of various structures. In this paper, we showed that the blurring mean-shift is a static SUP. Therefore our discussions on the strengths of SUP also apply to the blurring mean-shift.


\begin{classcode}91C20;62H30;68T10\end{classcode}

\end{abstract}

\section{Introduction}
Clustering analysis is a useful technique to discover patterns in
data. This technique has been widely applied to many disciplines for
partitioning data into groups. In the literature, a vast number of
clustering algorithms have been developed. The model-based methods
\citep{Banfield} make an assumption on the probabilistic
distribution of data, and the distance-based methods employ the
notion of ``distance" that represents the similarity between two
data points. Among the distance-based methods, two major types are
most commonly used. The first type is hierarchical clustering
\citep{Hartigan}, which clusters data into groups through a series
of agglomerative or divisive steps that operate on the similarity
measure between data points. The second type uses a clustering
criterion. Clustering results are obtained by optimizing the criterion.
The most popular clustering method of the second type is probably the k-means algorithm \citep{McQueen, Lloyd}.
A large number of follow-up articles have addressed to improve some weaknesses of the k-means \citep{Milligan,Tibshirani,Selim,Tseng}.
In addition, the fuzzy c-means algorithm \citep{fcm} was developed as a soft type of the k-means algorithm.
With the flexibility that each element in the data can belong to each of the clusters with probabilities,
the fuzzy c-means is more robust than the k-means in the presence of noise.

The Self-Updating Process (SUP) \citep{SUP2007} is a distance-based method for
clustering. The original idea was initiated as an extension of
the iteratively generated correlation matrices
\citep{McQuitty,CONCOR,GAP}, according to which data points are
gathered towards the left and right sides of an ellipse at each
iteration and eventually merge into two clusters. The sequence of
correlation matrices consequently produces an iterative process for
clustering, which has been implemented in the Generalized
Association Plots \citep{GAP,Wu}. Compared with the iteratively
generated correlation matrices, the self-updating process operates
on the sample space, not on the correlation space. It shows the
actual movements of data points around the sample space. Data points
continue updating their positions until the whole system reaches a
balanced therefore static condition, in which the clusters are formed.
It is as if the process describes how
data points perform self-clustering. We therefore named it
Self-Updating Process (SUP).

A similar iterative process that also operates on the sample space
is the mean-shift algorithm \citep{fukunaga}. It has non-blurring and blurring approaches.
Compared with the self-updating process in
which operators can be time-varying, both non-blurring and blurring
approaches use time-invariant operators.
Specific differences between the mean-shift and and the self-updating process are outlined in Section \ref{subsec:diff}.
The mean-shift algorithm made its first appearance for kernel density estimation by taking the
sample mean within a local region to estimate the gradient of a
density function. It was further extended and analyzed by Cheng
\citep{cheng}. Comaniciu successfully applied the non-blurring mean-shift algorithm to the
problem of image segmentation \citep{comaniciu}.
Since then the mean-shift algorithm has become well-known in the Computer Science community but not as familiar to the
Statistics community. As the implementation of the mean-shift algorithm requires a choice of the kernel function,
the Gaussian kernel is very often used in practice \citep{comaniciu,Carreira,Carreira2}.
There are other clustering algorithms that can be viewed as some version of the mean-shift algorithm.
For example, Cheng \citep{cheng} showed that the k-means algorithm is some limit of the non-blurring mean-shift algorithm.
Yang and Wu used a total similarity objective function to derive a similarity-based clustering method (SCM) \citep{scm},
which is a non-blurring mean-shift type clustering algorithm. Although the non-blurring mean-shift is more popular in image processing, Chen et al.
\citep{Chen3} reported that the blurring process is more robust and is often more efficient than the non-blurring process
in location estimation.

Through simulations and real data examples,
this paper discusses the strengths of the self-updating process in clustering
the following three types of data: (i) data with noise, (ii) data with a
large number of clusters, and (iii) unbalanced data.
Such data is often met in practice, but it is difficult to be analyzed. One
example is the Cry-EM image data, which has a high
level of noise and a large number of clusters. Chen et al.
\citep{gamma_sup} applied the $\gamma$-SUP, a variant of SUP that
minimizes the $\gamma$-divergence, to the analysis of Cryo-EM
images. The clustering results by $\gamma$-SUP correctly identified
all of the $128$ clusters of Cryo-EM images.
On the basis of the similarity between the blurring mean-shift and the self-updating process,
our discussions in this paper also apply to the blurring mean-shift type algorithms.
To our knowledge, such discussions on the strengths of the mean-shift type algorithms
have not been reported in the literature.

This paper is organized as follows. In Section \ref{sec: SUP} we
introduce the self-updating process in details. Section
\ref{sec: simu} presents simulation results for the aforementioned three types of data. Real data applications
are given in Section \ref{sec: example}. A discussion is provided in
the final section.

\section{The Self-Updating Process}
\label{sec: SUP}

\subsection{The idea}
Suppose there are $N$ elements to be clustered, and there are $p$
random variables representing elements' information. The data is a
$N \times p$ matrix. We can view this data as $N$ data points in a
$p$-dimensional space. When the updating process begins, data points
start to move. The movement of a point is determined by its
relationship with other data points. We can quantify the
relationships according to data points' information, using measures
such as the correlation, Euclidean distance, or other measures that
are relevant.

\subsection{Main algorithm}
\label{subsec: algorithm} The self-updating process is formulated as
follows.

\begin{enumerate}
\renewcommand{\labelenumi}{(\roman{enumi})}
\item $x_1^{(0)}, \ldots , x_N^{(0)}\in R^p$ are the original positions of data points to be clustered.
\item At time $t+1$, every point is updated to the following
new position:
\begin{equation} \label{eq:update}
x_i^{(t+1)}= \sum_{j=1}^N  \displaystyle
\frac{f_t(x_i^{(t)},x_j^{(t)}) } {\sum_{k=1}^N
f_t(x_i^{(t)},x_k^{(t)})}
                           x_j^{(t)},
\end{equation}
where $f_t$ is some function that measures the influence between two
data points at time $t$.
\item Repeat (ii) until every data point no longer moves.
\end{enumerate}
When two data points are closer, the influence between them should
be stronger. Therefore, we assign a larger value to $f_t$ when
$x_i^{(t)}$ and $x_j^{(t)}$ are closer. We interpret
$f_t(x_i^{(t)},x_j^{(t)})$ as the mutual influence between point $i$
and point $j$ at the $t$-th update. In plain words, equation
(\ref{eq:update}) states that the next position where point $i$ moves
to is determined by the influences it currently receives from all
data points, including from point $i$ itself. In statistical
terminology, $x_i^{(t+1)}$ is the weighted average of all
$x_i^{(t)}$'s, for $i \in \{1,...,N\}$.
Throughout this paper we take $f_t$ as a truncated exponential decay function of some distance $d$,
\begin{eqnarray}
&&f_t(x_i^{(t)},x_j^{(t)}) \nonumber\\
&=&  \left \{ \begin{array}{cl} \exp[-d(x_i^{(t)},x_j^{(t)}) / T(t)], & \quad d(x_i^{(t)},x_j^{(t)}) \leq r \\
0, & \quad d(x_i^{(t)},x_j^{(t)}) >r,
\end{array} \right.
\label{eq:f}
\end{eqnarray}
where $r$ and $T(t)$ are parameters, and $d(x_i^{(t)},x_j^{(t)})$ is the Euclidean distance between positions of point $i$ and point $j$ at the $t$-th update. We propose the use of exponential decay function because it is very often observed in nature. Other formulations of $f_t$'s can be considered and are discussed in the final section. Note that $f_t$ can change over iterations $t$'s. In such a case, we call the process {\it dynamic} SUP. When $f_t$ does not change with $t$, it is called {\it static}.

\subsection{A simple illustration}
\label{subsec: simple_example} 
\begin{figure}[th]
\centering
\includegraphics[width=0.75\textwidth]{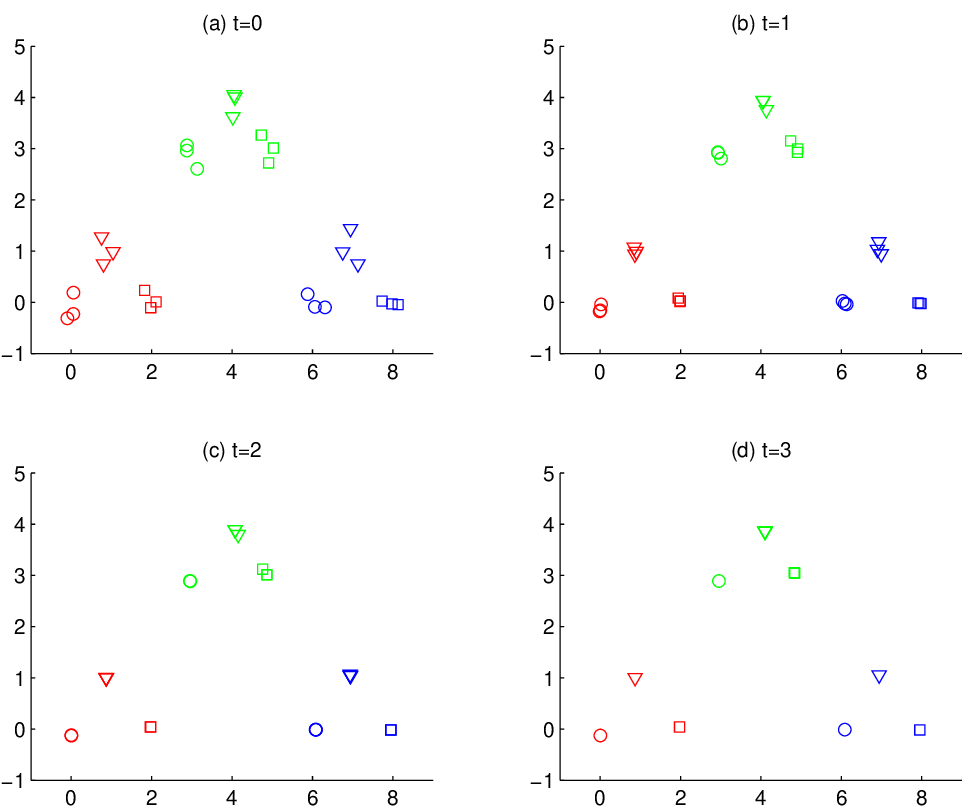}
\caption{SUP with r=0.9 and T=0.7} \label{fig:ex1-1}
\end{figure}

We present a simple example to
illustrate the self-updating process and the effects of parameters $r$
and $T$. Three data points from bivariate normal distributions
$BVN(\mu_k, I_2/25)$ were sampled for each $k \in \{1,...,9 \}$,
where $\mu_k \in \{$(0,0), (2,0), (1,1), (6,0), (8,0), (7,1), (3,3),
(5,3) and (4,4)$\}$. Figure \ref{fig:ex1-1}(a) plots a total of $27$ sampled data points.
With $r=0.9$ and $T=0.7$, Figure \ref{fig:ex1-1}(b)-(d) show the updated position of each data point at the first, second and third iteration.
The $27$ data points moved to nine positions at the third iteration and made no further movements. The nine positions
are the representative positions of the resulting nine clusters.


\begin{figure}[!h]
\centering
\includegraphics[width=0.75\textwidth]{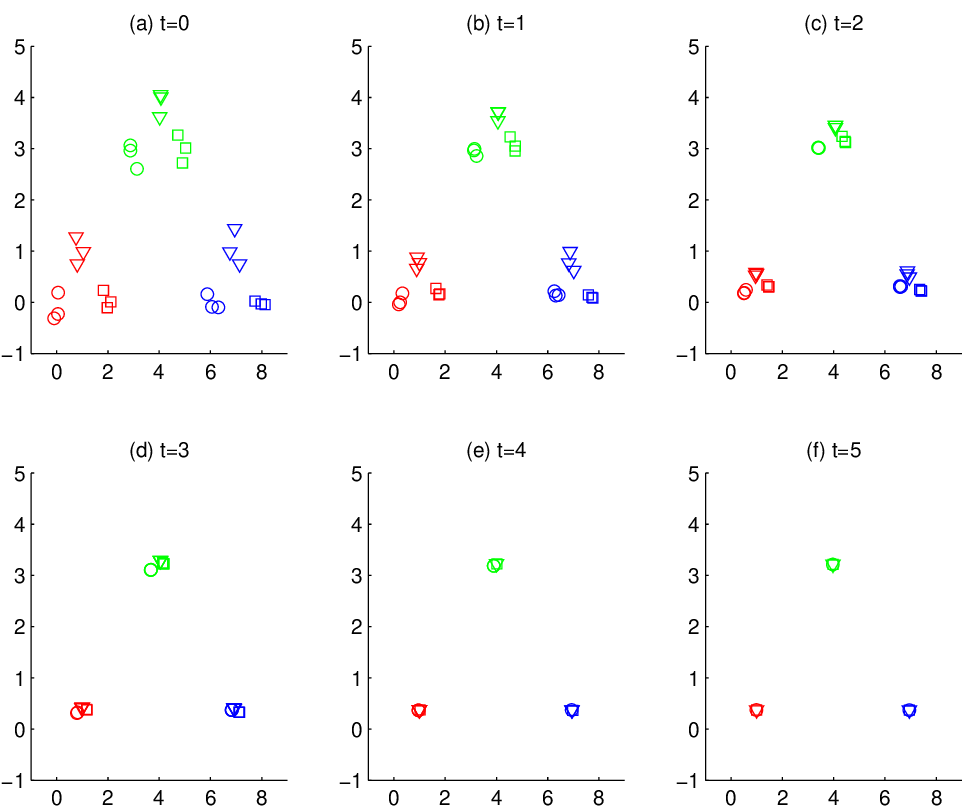}
\caption{SUP with r=3.5 and T=0.7} \label{fig:ex1-2}
\end{figure}

\subsubsection{The parameter $r$}
Figures \ref{fig:ex1-2}(a)-(f) present the updating process and
the final clustering result when $r$ was increased to $3.5$. In the end of the process, data points moved to three
positions instead of nine. The difference between Figure \ref{fig:ex1-1}(d) and Figure \ref{fig:ex1-2}(f) explains the effect of $r$.
In Figure \ref{fig:ex1-1}(d), a choice of $r=0.7$ forced each data point to be influenced only by those within $0.7$ units.
Squares, circles and triangles of the same colors had no influence on one another, therefore eventually moved to different positions.
We interpret $r$ as the {\it range of influence}. The use of a small $r$ value generally produces clusters of compact
sizes. Without the use of $r$, or equivalently, when $r$ is
infinite, Corollary \ref{cor:f>0} in the next section proves that
all data points eventually move to one position when $f$ is a strictly
positive function.

\begin{figure}[htb]
\centering
\includegraphics[width=\textwidth]{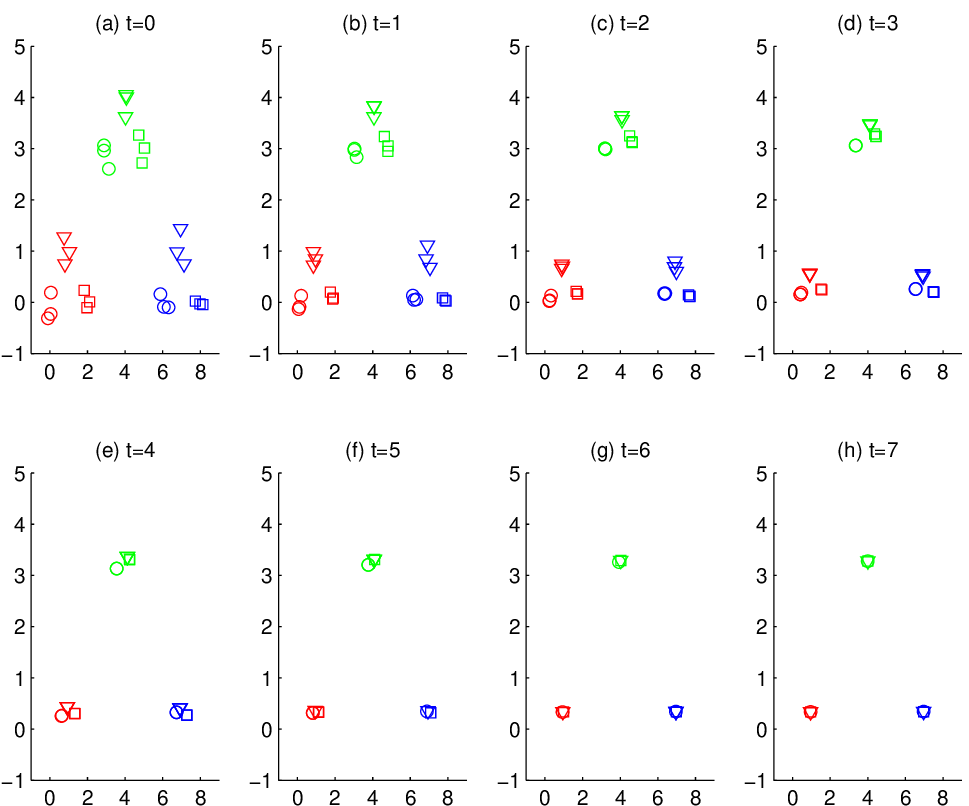}
\caption{A graphical presentation of SUP with r=3.5 and T=0.5}
\label{fig:ex1-3}
\end{figure}

\subsubsection{The parameter $T$}
\label{subsec:t}
Figures \ref{fig:ex1-3}(a)-(h) present the
updating process and the final clustering result when $T$ was decreased to $0.5$ and when $r$ was $3.5$.
A comparison between Figure \ref{fig:ex1-2} and Figure \ref{fig:ex1-3} shows
that data points moved at a slower rate when $T$ was smaller.
This observation can be explained mathematically from (\ref{eq:f}): When $T$ is
small, $f(u,u)=1$ is much larger than $f(u,v)$ for every $v \ne u$.
Data point $u$ therefore hardly moves as the influence from itself
totally dominates. Similarly, it is can be explained that data points move faster when $T$ is larger. If we consider data
points as particles in a statistical mechanical system, then
parameter $T$ can be interpreted as {\it temperature}. This
temperature parameter determines the speed of the updating process.

\subsection{The effect of parameters and the proposed parameter values}\label{sec:ps}

\subsubsection{The influential range $r$}
When there is a training set, cross-validation is a standard way to
estimate the value of $r$. However, in practice we rarely have
additional data sets to learn the parameter values. In the following
we present a simple data-driven approach. The distribution of the
pairwise distance provides useful information on the structure of
data.

We begin with a simple situation when there are only two clusters in
the data. Confined to the two-cluster structure, the pairwise
distances of pairs that contain one point from each cluster should
not differ much, meaning that the estimated probability density
function of the pairwise distance has a large probability mass in
the range of the between-cluster distances. Similarly, the pairwise
distances of pairs that contain both points from the same cluster
should not differ much. There should also be a large probability
mass in the range of the within-cluster distances. To select an
influential range $r$ that produces clusters retaining the original
structure of two clusters, we should avoid the distances at peak
regions that are likely to be the between- or within-cluster
distances. We propose to select the distance at sharp valley
regions, because this valley selection can reduce the chance that
the updating process distorts the data structure, as the number of
pairs that are influenced by this selection is small. The same
reasoning applies to data of more than two clusters.

\begin{figure}[h]
\centering
\includegraphics[height=1.2in,width=0.5\textwidth]{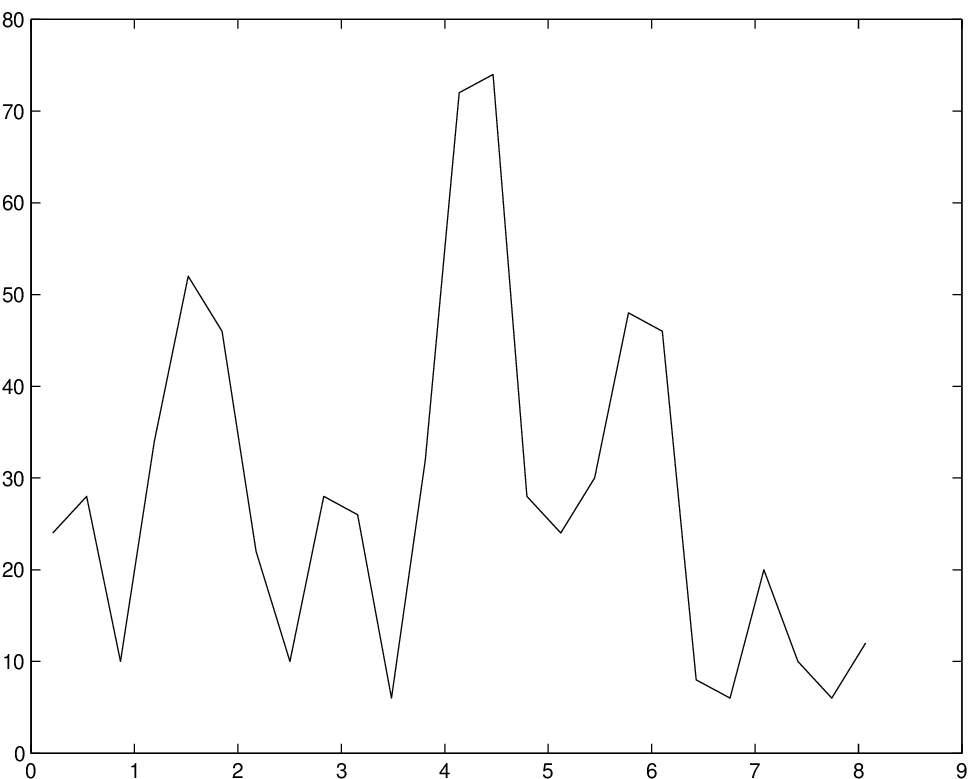}
\caption{The frequency polygon of the pairwise distances.}
\label{fig:pl-1}
\end{figure}

Take the data presented in Figure \ref{fig:ex1-1}(a) for
example. We use frequency polygon to approximate the probability
density function of the pairwise distance. Figure \ref{fig:pl-1}
presents the frequency polygon, in which sharp valleys occur at
around $0.9$, $2.5$, $3.5$, $5.1$, $6.8$ and $7.7$. We showed in
Section \ref{subsec: simple_example} that the use of $r=0.9$
produced three clusters and that of $r=3.5$ produced nine clusters.
For the rest of the valleys, $r=2.5$ produced an identical
clustering result as $r=3.5$, while $r=5.1$, $6.8$ and $7.7$ moved
all data points into one single cluster.

When data does not show clear patterns of clusters or when the noise
level in data is substantial, the frequency polygon may not have
peaks or valleys. In such situations, we can try different values of $r$ by taking percentiles of the pairwise distances.

\subsubsection{The temperature $T$}
\label{sec:temperature}
In Section $\ref{subsec:t}$, we showed that the temperature $T$ influences the
updating speed of SUP. Therefore, one may want to choose a large $T$ value to reduce the computation time.
This is, however, not necessarily a good choice. For data with certain structures,
the updating speed is also likely to influence the clustering results.

In the following we describe two examples of extreme $T$ values.
When $T$ is set to be very large, $f_t$ is close to one for every pair of $x_i^{(t)}$
and $x_j^{(t)}$. Namely, each data point receives nearly the same amount of influences from the other data points,
suggesting that the updating process utilizes the information on the overall structure of the data.
When $T$ is set to be close to zero, each data point receives comparatively negligible influences from the other data points,
except from those that are the closest.
That is, the updating process with such a small $T$ value utilizes the information on the most local structure of the data.
To summarize, the temperature $T$ that represents the speed of the updating process also determines the relative emphasis on the global versus the local structures.
Different levels of emphasis mostly influence the data points located near the boundaries of the clusters.
From the descriptions above, the effect of temperature $T$ is summarized as follows.
When the structures of the clusters in data are simple,
such as well separated clusters and clusters of similar sizes and shapes,
the boundary points of one cluster are often located far enough from those of another cluster.
In this case, it is easier to select a $r$ value so that there is no mutual influences between boundary points of different clusters.
The temperature $T$ therefore only determines the updating speed.
The illustration presented in Section $\ref{subsec: simple_example}$ is such an example.
When the structures of the clusters in data are complex,
the selection of a proper $r$ value is difficult.
The temperature $T$ therefore influences the boundary data points and consequently the final clustering results.

To select a temperature $T$, we consider the following.
Suppose that data point $i$ is $r-\delta$ and $r+\delta$ units away from data points $j$ and $k$,
respectively, where $r$ is the influential range. When
$\delta$ is small, $i$ is about $r$ units away from both $j$ and
$k$. It is reasonable to assume that $i$ receives approximately the
same amount of influence from $j$ and $k$. According to
(\ref{eq:f}), however, the actual influences that $i$ receives from
$j$ and $k$ are $\exp[-(r+\delta)/T]$ and zero, respectively.
We therefore propose to use a small enough $T$ value so that $\exp(-r/T)$ is
close to zero. For static temperature, throughout this paper we use $T=r/5$, which makes
$\exp(-r/T)=0.0063$. Our experiments showed that the use of $T=r/5$
very often produced good clustering results within a reasonable
computing time.

When data contains different sizes and shapes of clusters,
the use of local structures of data is usually more capable of correctly identifying clusters of various structures.
The idea is to capture more of the local structures by allowing data points to move at a low temperature
in early iterations. Afterwards, the temperature is gradually increased with time to accelerate the
updating speed. For dynamic temperature, we need to select the {\it initial temperature} $T_0$ and the {\it heating rate} $s$,
where $T_0$ is certainly smaller than $r/5$.
Throughout this paper we present results from using $T(t)=T_0+st$,
where $T_0=r/20$ and $s=r/50$. In this heating scheme,
the temperature increases linearly with time and exceeds $r/5$ after the seventh iteration.
Different values of $T_0$ and $s$ can be considered according to the running time and the emphasis
on the local versus global properties. However, our experiments showed that the dynamic SUP is relatively more stable than the static SUP:
the use of different temperatures $T_0$ and $s$ has less influence on the structures of the final clustering results.
The simulation results in Section \ref{sec: simu} also support this statement.

\subsection{Convergence}
The self-updating process stops when all the data points no longer
move. This is called the convergence. The convergence of the
self-updating process depends on the function $f$. Chen
\citep{Chen2} showed that {\it PDD} (positive and decreasing with
respect to distance) is a sufficient condition to ensure the
convergence.

\begin{definition}
A function $f$ is {\it PDD} (positive and decreasing with respect to
distance), if
\begin{enumerate}
\renewcommand{\labelenumi}{(\roman{enumi})}
\item $0 \leq f(u,v) \leq$ 1, and $f(u,v)=1$ only when $u=v$.
\item $f(u,v)$ depends only on $\|u-v\|$, the distance between u and v.
\item $f(u,v)$ is decreasing with respect to $\|u-v\|$,
\end{enumerate}
\end{definition}

The PDD condition specifies $f$ to have the following properties.
Condition (i) states the non-negativity of $f$. This condition
excludes the situation that data points $u$ and $v$ may repel each
other. Although in practice $f(u,v)$ may be negative, data points
move further apart and eventually may result in the divergence of
the updating process. This is the reason we require $f(u,v)$ to be
non-negative. Generally $f(u,u)$ can be defined as any positive
number. For simplicity it is normalized to be one. Condition (ii)
states that the influence between $u$ and $v$ is solely determined
by the distance between them, meaning that $f(u_1,v_1)=f(u_2,v_2)$
whenever $\|u_1-v_1\|=\|u_2-v_2\|$. Condition (iii) states that the
influence between $u$ and $v$ is larger whenever $u$ and $v$ are
closer.

The following Theorem \ref{thm:main} guarantees the convergence of
SUP when $f_t$ satisfies the PDD condition for each $t$. The proof
of the case that $f_t$'s are time-invariant can be found in
\citep{Chen2}. The proof for the case of time-varying functions
$f_t$'s is essentially the same.

\renewcommand{\labelenumi}{(\roman{enumi})}
\newtheorem{thm}{Theorem}

\begin{thm}\label{thm:main}
If the function $f_t$ in (\ref{eq:update}) is PDD for each $t$,
there exists $\{x_1, \ldots, x_N\}$, such that
\[
\lim_{t \to \infty} x_i^{(t)} = x_i \quad\quad \forall i.
\]
\end{thm}

The following corollary further identifies $f_t$'s that produce
trivial clustering results: If $f_t$ are PDD and strictly positive,
all data points converge to one single position.

\begin{corollary}\label{cor:f>0}
Let $r_M$ be the maximum pairwise distance between any two data
points. If $f_t$'s are PDD with $f_t(a,b)>0$ whenever $||a-b||\leq
r_M$, there exists $c$, such that
\[
\lim_{t \to \infty}x_i^{(t)}=c \quad\quad \forall i.
\]
\end{corollary}

Recall that we introduced the parameter $r$ in (\ref{eq:update}).
Corollary \ref{cor:f>0} shows the necessity of the use of $r$ in the
specification of any $f_t$ function.

\subsection{The blurring mean-shift is a static SUP}
\label{subsec:diff} Although the mean-shift and the self-updating
process were developed independently, their mathematical forms
display great similarity. The original mean-shift \citep{fukunaga}
uses the operator:
\begin{equation}
\label{eq:update_meanshift}
x_i^{(t+1)}=x_i^{(t)}+a \nabla_x \ln (p(x_i^{(t)})),
\end{equation}
where $a$ is a positive constant to ensure the convergence, $\nabla$
is the gradient, $\ln$ denotes the natural logarithm, and $p$ is the density function. Cheng \citep{cheng}
generalized (\ref{eq:update_meanshift}) to the following form:
\begin{equation} \label{eq:update_cheng}
x_i^{(t+1)}= \sum_{j=1}^N  \displaystyle
\frac{K(x_i^{(t)}-x_j) } {\sum_{k=1}^N K(x_i^{(t)}-x_k)}
                           x_j,
\end{equation}
where $x_j$'s represent the positions of data points. When the
positions at the current iteration $x_j^{(t)}$'s are used, the algorithm is called the
{\it blurring} mean-shift. In contrast, we use {\it non-blurring} to refer to the mean-shift that uses
the initial positions $x_j^{(0)}$'s.

The self-updating process (\ref{eq:update}) that uses the current positions is closer to the {\it blurring} mean-shift algorithm.
In the following we outline the specific differences between the two.

\begin{enumerate}
\item $K$ is originally a flat kernel. It is an indicator function representing
whether $x_i^{(t)}-x_j$ is less than some threshold
value. Cheng generalized $K$ to be any kernel function.
In most of the mean-shift applications, truncated Gaussian kernels are often considered.
SUP is less restricted in the sense that the integrability of $f_t$ is not required.
\item The mean-shift uses the same kernel $K$ over iterations. SUP uses $f_t$ that can change over
iterations.
\end{enumerate}

From the above, we see that the blurring mean-shift is a static SUP that uses a fixed $f_t$ function through iterations.
In Section \ref{sec:temperature} we described the advantage of using the dynamic temperature in plan words.
The following Section \ref{subsec: large clusters} and \ref{subsec: simu_crowded} in particular address the
different impacts on clustering performance between the static (the blurring mean-shift) and the dynamic SUP.


\section{Simulation and Comparison}
\label{sec: simu} This section shows the strengths of the
self-updating process in clustering the following three types of
data: (i) data with noise, (ii) data with a large number of
clusters, and (iii) unbalanced data. In each simulation example,
we compare clustering results of the k-means,
the fuzzy c-means, the non-blurring mean-shift, the static SUP,
and the dynamic SUP.
We replace the current positions $x_j^{(t)}$'s in ($\ref{eq:update}$) with the initial positions $x_j^{(0)}$'s
and take $r$ in ($\ref{eq:f}$) as infinity and $T(t)$ as fixed to represent the non-blurring mean-shift type algorithms.

Throughout the three simulation examples in this section, we use the the following temperatures.
For the non-blurring mean-shift, we experimented with several $T$ values and selected the best performance.
For the static and the dynamic SUP, $T=r/5$ and $T(t)=r/20+(r/50)t$, respectively.

\subsection{Data with noise}
\label{subsec: simu_noise} Data that contains noise is very often
present in practice, such as the gene expressions data and the image
data. Clustering algorithms sometimes fail to produce
reasonable results for data with noise, because scattered points of
noise can very often obscure the structure of data, therefore
make it difficult for algorithms to discover patterns.

\begin{figure}[h]
\centering
\includegraphics[width=0.75\textwidth]{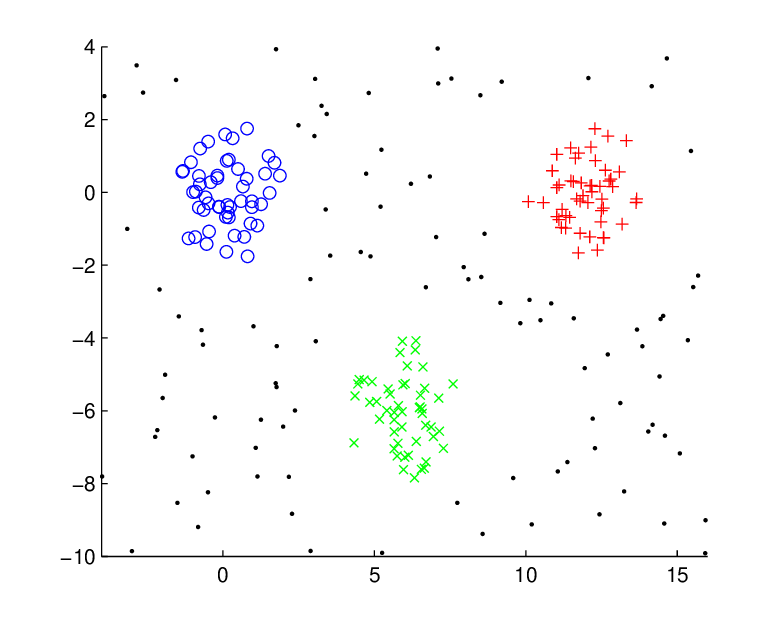}
\caption{Three groups (circles, x-marks and pluses) and noises
(dots)} \label{fig:ex2}
\end{figure}

We used Tseng and Wong's example \citep{Tseng} to compare the performance of the following algorithms on data with noise:
the k-means, the fuzzy c-means and the mean-shift type algorithms, including SUP.
The example includes three clusters and a number of scattered points.
Data points in the three clusters were sampled from standard normal distributions
centered at (-6,0), (6,0) and (0 6), respectively. Each point was restricted within two standard deviations to its
center. The scattered points representing noise were sampled
uniformly from [-12, 12] $\times$ [-6, 12], but not within three
standard deviations to any of the three centers. Each simulated data contained $50$ points from each of the
three clusters and $n$ scattered points, where $n$ can be $10$,
$50$, $100$ or $200$, representing varying degrees of noise. Figure \ref{fig:ex2} shows one simulated
data: $50$ points from each of the three clusters were denoted by
circles, x-marks and pluses, and $50$ scattered points of noise were denoted by
dots.

The selection of parameter values are as follows.
For the k-means algorithm, we present results from using random initials of one set, random
initials of $100$ sets, and the initials proposed by Tseng and Wong using Hierarchical clustering with single
and complete linkages by taking p=$1$, $3$, and $6$, respectively.
The latter two approaches were used to solve the problem of local minimum for the k-means.
When the random initials of $100$ sets were used,
only the clustering result from the set of initials that achieved the
minimum sum of the within-cluster variations is presented for comparison.
For the static and dynamic SUP, the value of $r$ was selected
automatically at a time for each simulated data according to the frequency polygon of the pairwise distances.


Since the purpose of this example is to correctly distinguish between the three clusters in the presence of noise,
we only compare the clustering results of the $150$ non-noise data points.
Table \ref{tab:comp} presents the number of incorrect runs out of the $100,000$ runs of simulations for each level of noise,
in which ''incorrect" means that at least one of the $150$ non-noise data points was clustered incorrectly.
Table \ref{tab:comp} shows that both the blurring and non-blurring mean-shift type algorithms are robust against noise.
The fuzzy c-means also demonstrates to be robust.

We further compared the running time in seconds for one
run of simulation. Results are presented in Table \ref{tab:cpu}.
The fuzzy c-means shows great promise in both accuracy and computation efficiency when noise is present in the data.
The non-blurring mean-shift and the self-updating process are competitive in accuracy with a reasonable computation efficiency.

\begin{table}[h]
\caption{The numbers of incorrect runs in 100,000 runs of
simulations}\label{tab:comp}
\centering
\begin{tabular}{|l|l|l|r|r|r|r|}\hline
\multicolumn{3}{|l|}{Number of noise}&10&50&100&200\\
\hline & \multicolumn{2}{|l|}{One random initial set} &9746&1248&1179&1481 \\
\cline{2-7} & \multicolumn{2}{|l|}{100 random initial sets} &0&0&0&731 \\
\cline{2-7}&&$p$=1 &6934&1453&1229&1165 \\
\cline{3-7} &\raisebox{1.5ex}[0pt]{Single} &$p$=3 &41&3161&2155&1183\\
\cline{3-7} K-means& \raisebox{1.5ex}[0pt]{Linkage} &$p$=6
&0&534&956&944\\
\cline{2-7} &&$p$=1 &1860&1294&1054&1253 \\
\cline{3-7} &\raisebox{1.5ex}[0pt]{Complete} &$p$=3 &9&0&0&483\\
\cline{3-7} & \raisebox{1.5ex}[0pt]{Linkage} &$p$=6 &4517&4&0&363 \\
\hline \multicolumn{3}{|l|}{Fuzzy c-means with one random initial set} &0&0&0&0\\
\hline \multicolumn{3}{|l|}{Non-blurring mean-shift} &0&0&0&0\\
\hline \multicolumn{3}{|l|}{Static SUP (Blurring mean-shift)} &0&0&0&0\\
\hline \multicolumn{3}{|l|}{Dynamic SUP} &0&0&0&1\\\hline
\end{tabular}
\end{table}

\begin{table}
\caption{\label{tab:cpu}CPU time per run in seconds} \centering
\begin{tabular}{|l|l|r|r|r|r|}
\hline
\multicolumn{2}{|l|}{Number of noise}&10&50&100&200\\
\hline & One random initial set &0.003&0.003&0.004&0.004 \\
\cline{2-6} \raisebox{1.5ex}[0pt]{K-means}& 100 random initial sets &0.149&0.171&0.223&0.246 \\
\hline \multicolumn{2}{|l|}{Fuzzy c-means with one random initial set }&0.003&0.004&0.005&0.012\\
\hline \multicolumn{2}{|l|}{Non-blurring mean-shift }&0.088&0.434&0.798&1.738\\
\hline \multicolumn{2}{|l|}{Static SUP (Blurring mean-shift) }&0.020&0.056&0.092&0.248\\
\hline \multicolumn{2}{|l|}{Dynamic SUP }
&0.026&0.053&0.081&0.199\\\hline

\end{tabular}
\end{table}

\subsection{Large number of clusters}\label{subsec: large clusters}
For data with a large number of clusters, it is often difficult for
most clustering methods to correctly identify every cluster. This is
especially true for methods that require a set of initial values,
such as the k-means algorithm that relies on a good initial
assignment of cluster centers. When the number of cluster is large,
the chance that each assigned initial center is close to a true
center is small. As a result, clusters without any assigned initial
center are likely to be absorbed into other clusters, and clusters
within which multiple initial centers are assigned are likely to be
broken down to multiple clusters. In comparison with the k-means,
the blurring and the non-blurring mean-shift type algorithms do not require initial values to begin
the updating process. The process moves each data point locally
towards where most neighboring data points are, making it possible
to capture every high-density region of data points.

\begin{figure}[htb]
\centering
\includegraphics[width=\textwidth]{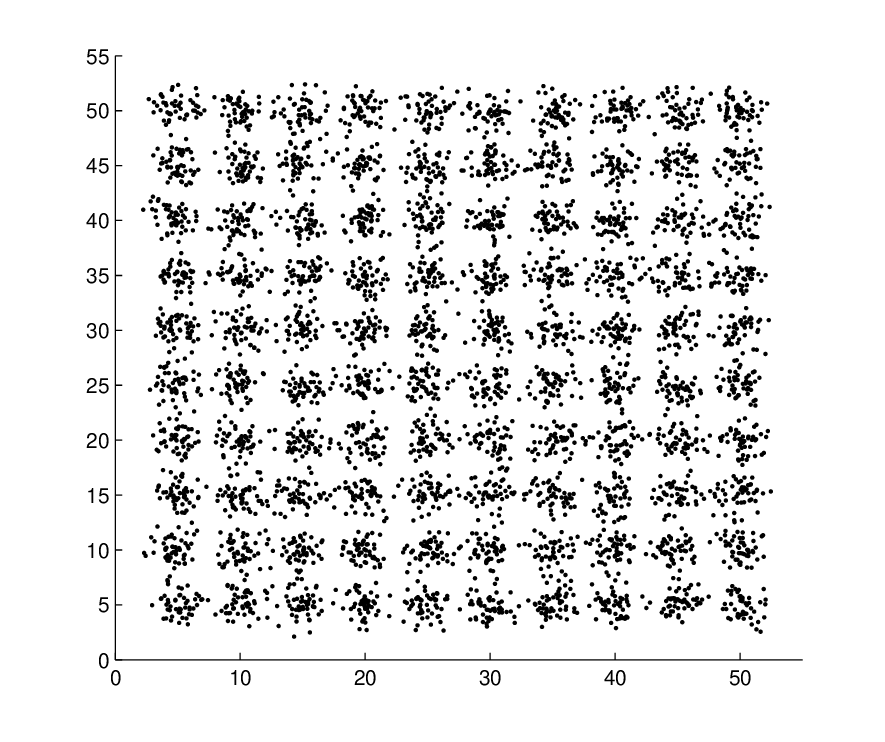}
\caption{Example of simulated data set of 100 clusters}
\label{fig:1010}
\end{figure}

This simulation example considers data with one hundred clusters.
For each cluster, $50$ points were sampled from standard normal
distributions centered at $(5\times i, 5 \times j)$, where $i,j \in
\{1,\ldots,10\}$. Each point in a cluster was restricted to within
three standard deviations from its center. Figure \ref{fig:1010}
shows one simulated data. Note that the diameter of each cluster is
six and the distance between centers of any two clusters is five,
meaning that clusters are likely to overlap. For data points located
in overlapping regions, the number of mis-clustering errors is not
able to properly assess the clustering performance.
Therefore, we consider the sum of within-cluster
variations as a more appropriate metric for the evaluation of
clustering results in this simulation example .

We simulated $1000$ sets of data to compare the performance of clustering algorithms.
For the k-means and the fuzzy c-means, we used 100
sets of random initials and selected the one that produced the
smallest sum of within-cluster variations. For both static and
dynamic SUP, we used $r=3.6$ so that the number of clusters did not
exceed 100. This choice of $r$ was in order not to make an unfair
comparison to the k-means and the fuzzy c-means, because the sum of within-cluster
variations decreases with the increasing number of clusters. Table
\ref{tab:comp2} presents the clustering results from $1000$
simulated data, summarizing the means and the standard deviations of
the sum of within-cluster variations. These results show that the
self-updating process produced smaller sums of variations in
reasonable running time.

\begin{table}[h]
\caption{The means and the standard deviations in parentheses of the
sum of within cluster variations and the cpu time per run in
seconds}\label{tab:comp2} \centering
\begin{tabular}{|l|c|c|}
\hline & Sum of within-cluster variations & CPU time per run\\
\hline
 K-means with 100 initial sets& 11681 (363)& 27.84 (1.80) \\
 \hline
  Fuzzy c-means with 100 initial sets& 18442 (467)& 584.58 (21.50) \\
 \hline
 Non-blurring mean shift & 9382 (241) & 62.53 (17.84)\\
\hline
 Static SUP (Blurring mean shift)& 9185 (118) & 18.24 (1.94) \\
 \hline
Dynamic SUP & 9219 (128) & 25.07 (3.69) \\
\hline

\end{tabular}
\end{table}

\begin{figure}[htb]
\centering
\includegraphics[width=\textwidth]{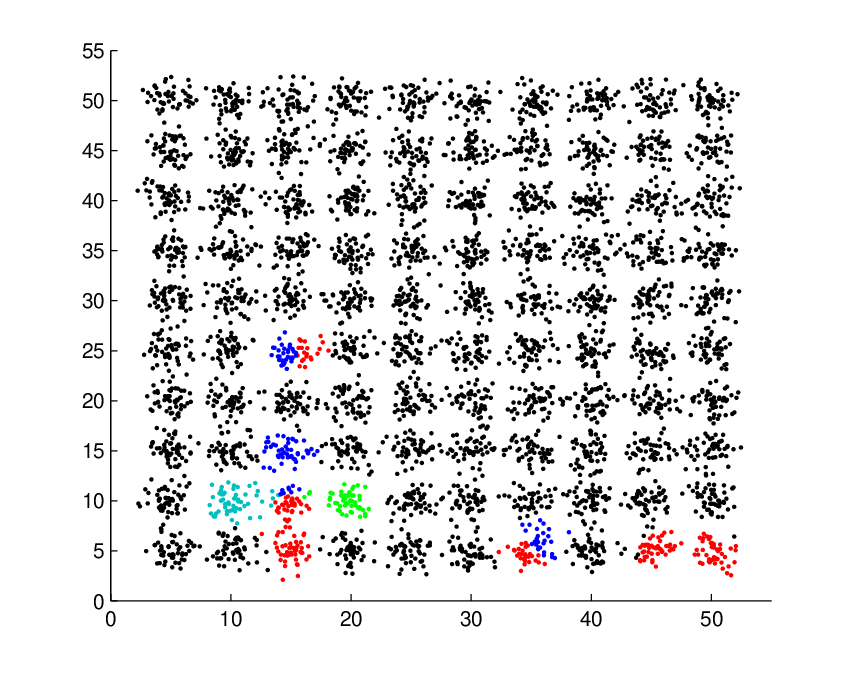}
\caption{Clustering results by the k-means using $100,000$ sets of
random initials} \label{fig:kmeans10x10}
\end{figure}

We increased the number of random initial sets to improve the
performance of the k-means and the fuzzy c-means. Our simulations showed that both algorithms
still failed to find the minimum sum of within-cluster
variations even when the number of random initial sets increased to
$100,000$. Illustrating using one simulated data, Figure
\ref{fig:kmeans10x10} shows the best solution from $100,000$ sets of
random initials by the k-means. The black color is used to represent
clusters that were correctly identified by the k-means, and the
other colors are to represent mis-clustered data points: points that
were clustered in the same group were displayed by the same color.
This color presentation makes it easier to see how the k-means
falsely split and merged some of the true clusters.

In contrast to its great performance in the previous example when noise is present in data,
the fuzzy c-means in this example performs poorly in both the clustering results and the computation time.
First, the computation complexity of the fuzzy c-means
increases linearly with the number of clusters.
The running time of the fuzzy c-means is therefore much longer
than that of the k-means when data has a large number of clusters.
Second, the robustness against noise presented in the previous example is achieved by
assigning the noise data points with almost uniform probabilities of belonging to each of the three clusters.
Such an assignment minimizes the influences of the noise data points on the clustering results.
When the number of clusters is large so that good initial centers are difficult to be selected,
data points located far from the initial centers are likely to be mistaken as the noise data points.
Consequently, the fuzzy c-means cannot correctly identify the true clusters without an initial center nearby.

\begin{figure}[htb]
\centering
\includegraphics[width=\textwidth]{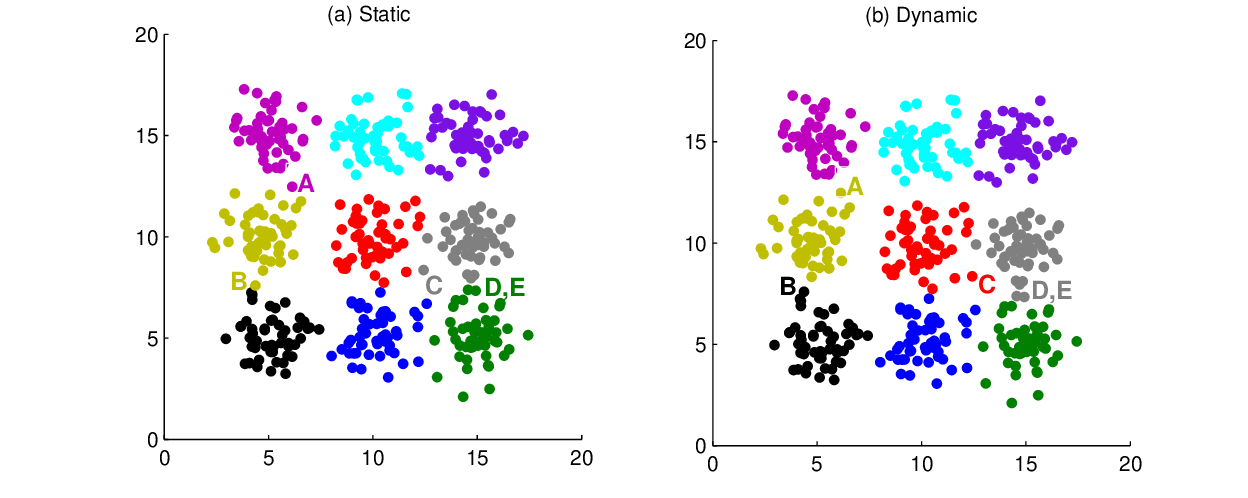}
\caption{Part of the clustering results by the static and the dynamic
SUP}\label{fig:sup_comp}
\end{figure}

The simulation results showed that the
non-blurring mean-shift, the static SUP and the dynamic SUP always correctly
identified the global structure of the $10 \times 10$ clusters.
Differences only existed on the boundary data points.
The non-blurring mean-shift reported a total within-cluster variations approximately $2\%$ higher than
that of the static and dynamic SUP. It produced a relatively more mixed structure of boundary data points
which were identified as from different clusters.
In contrast, the static and the dynamic SUP produced a cleaner structure of the clustering results on the boundary points.
Labeling the clustering results by letters A to E,
Figure \ref{fig:sup_comp}(a) and Figure \ref{fig:sup_comp}(b) compare the two temperature schemes.
The two figures show that the boundary data points are more influenced by their nearest neighbors in a dynamic SUP than in a static SUP.
This observation corresponds to our previous descriptions on the effect of the temperature:
The dynamic SUP utilizes more of the local structures of the data than the static SUP.

\subsection{Unbalanced data}
\label{subsec: simu_crowded}

When clusters are well separated and the sizes of the clusters are
similar, most of the clustering algorithms are able to provide good results.
When clusters are closely located and their sizes vary, clustering
becomes a challenging task.

\begin{figure}[htb]
\centering
\includegraphics[width=\textwidth]{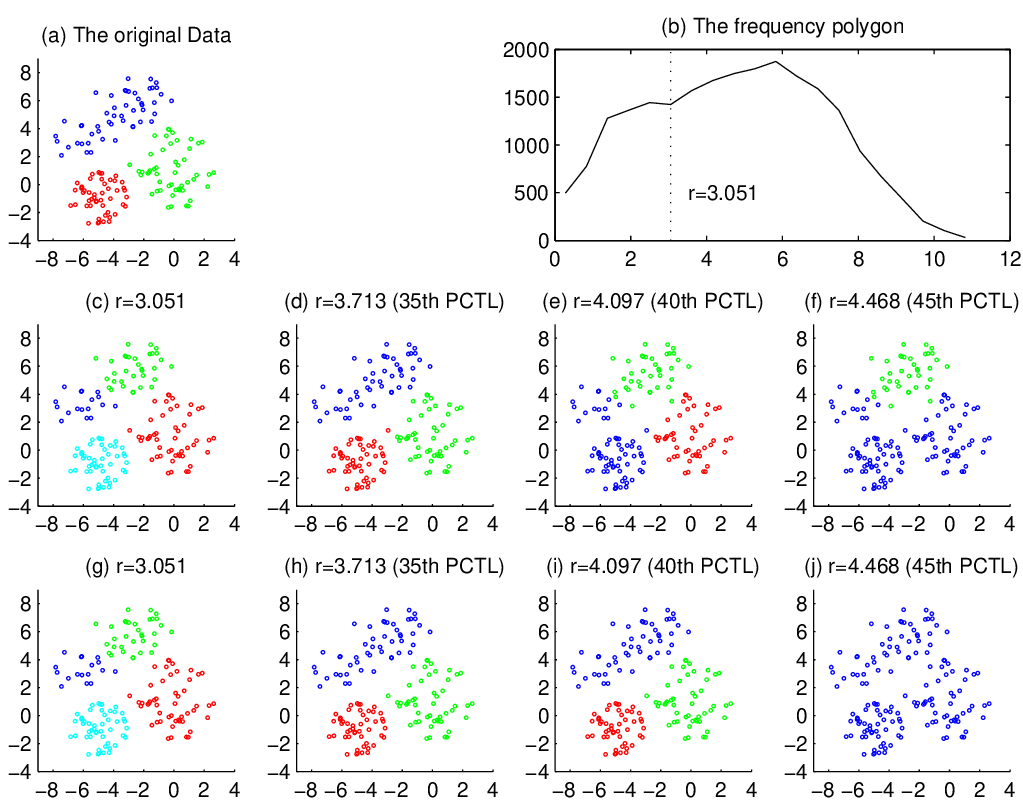}
\caption{(a) The example data. (b) The selection of $r$ value.
(c)-(f) Clustering results by the static SUP using various $r$ values.
(g)-(j) Clustering result by the dynamic SUP using various $r$ values.
Data points that were clustered into the same group are displayed by
the same color.} \label{fig:ex3-1}
\end{figure}

We sampled $30$ points for each of the three clusters from bivariate
normal distributions centered at $(-4, 5)$, (-5, -1) and $(0, 1)$
with $(\sigma_x, \sigma_y, \rho)$ as $(5, 2, 0)$, $(2, 2, 0)$ and
$(3, 3, 0)$, respectively. Each point was restricted to within one
standard deviation from its center. The data points in the first
cluster were further rotated counter-clockwise with respect to its
center $(-4, 5)$. This created the structure of an inclined ellipse.
Figure \ref{fig:ex3-1}(a) shows one simulated data. Data points
from the three clusters were colored in navy, red and green,
respectively. In the following we use this simulated data to compare the performances of the clustering algorithms.

\begin{figure}[htb]
\centering
\includegraphics[width=0.4\textwidth]{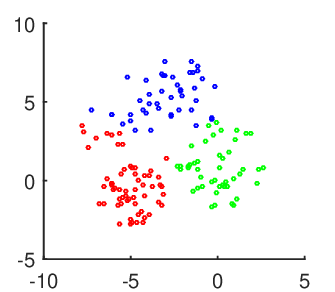}
\caption{Using the true centers are the initials,
the k-means and the fuzzy c-means provided the same clustering results.} \label{fig:ex3_kmeans}
\end{figure}

\begin{figure}[htb]
\centering
\includegraphics[width=\textwidth]{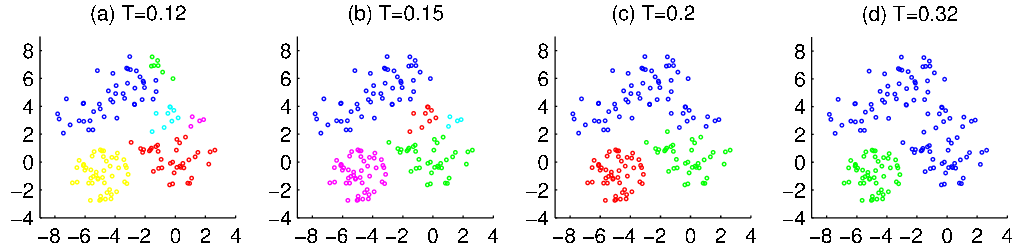}
\caption{The clustering results by non-blurring mean shift with
weight function $f=\exp{(-d/T)}$.} \label{fig:ex3_nb}
\end{figure}

Figure $\ref{fig:ex3_kmeans}$ presents the clustering results by the
k-means and the fuzzy c-means. Both algorithms failed to identify the structure of the
data, even though the ''true" centers were taken as the initials.
Figure \ref{fig:ex3_nb} presents the clustering results by the non-blurring mean-shift.
The temperature $T$ was selected to show the change of the structures of the clustering results as the number of clusters decreases.
These results show that the non-blurring mean-shift was able to correctly distinguish between the red and the navy clusters.
However, Figure \ref{fig:ex3_nb}(c) shows that some of the data points located near the boundary
of the green cluster were mis-identified.

For the static and the dynamic SUP, the selection of the $r$ value is as follows.
Figure \ref{fig:ex3-1}(b) is the frequency polygon calculated from the simulated data in Figure \ref{fig:ex3-1}(a).
There does not exist a sharp valley in this frequency polygon, because clusters in this example are closely located.
We only see a flat valley at around $3.051$. This $r$ value produced four clusters.
Figures \ref{fig:ex3-1}(c) and \ref{fig:ex3-1}(g) show the clustering results by the static and the dynamic SUP, respectively.

We increase the value of $r$ to reduce the number of clusters.
Figures \ref{fig:ex3-1}(d)-(f) and \ref{fig:ex3-1}(h)-(j) show the clustering results by the static and the dynamic SUP, respectively,
taking $r$ value as the $35$th, $40$th, and $45$th percentiles of the pairwise distances.
The results show that the static SUP made few mistakes on the data points located near the boundaries of the clusters.
In addition, Figure \ref{fig:ex3-1}(e) and Figure \ref{fig:ex3-1}(f) show the consequence of using an inappropriate $r$ value:
the structures of the data are falsely identified.
In contrast, Figure \ref{fig:ex3-1}(h) and Figure \ref{fig:ex3-1}(i) show that the dynamic SUP
correctly identified the structures of the data regardless of the $r$ values.

We described in Section \ref{sec:temperature} that the dynamic SUP utilizes more of the local structures of the data by
imposing a slow updating speed at early iterations. By showing that the dynamic SUP provided better clustering results than other algorithms,
this simulation example demonstrates the following:
When data contains closely located clusters which also have varying sizes, the ability of a clustering algorithm to utilize the local
structures for clustering becomes more important.

\section{Application}
\label{sec: example}
Section \ref{sec: simu} showed the strengths of the self-updating process in clustering
the three types of challenging data. In particular, results showed that the dynamic SUP was more capable to handle
data with various structures. Therefore, in this section we select the dynamic SUP to perform clustering analysis on the following real data applications.

\subsection{Seeds Data}
\label{subsec:seeds}
\begin{figure}[htb]
\centering
\includegraphics[width=\textwidth]{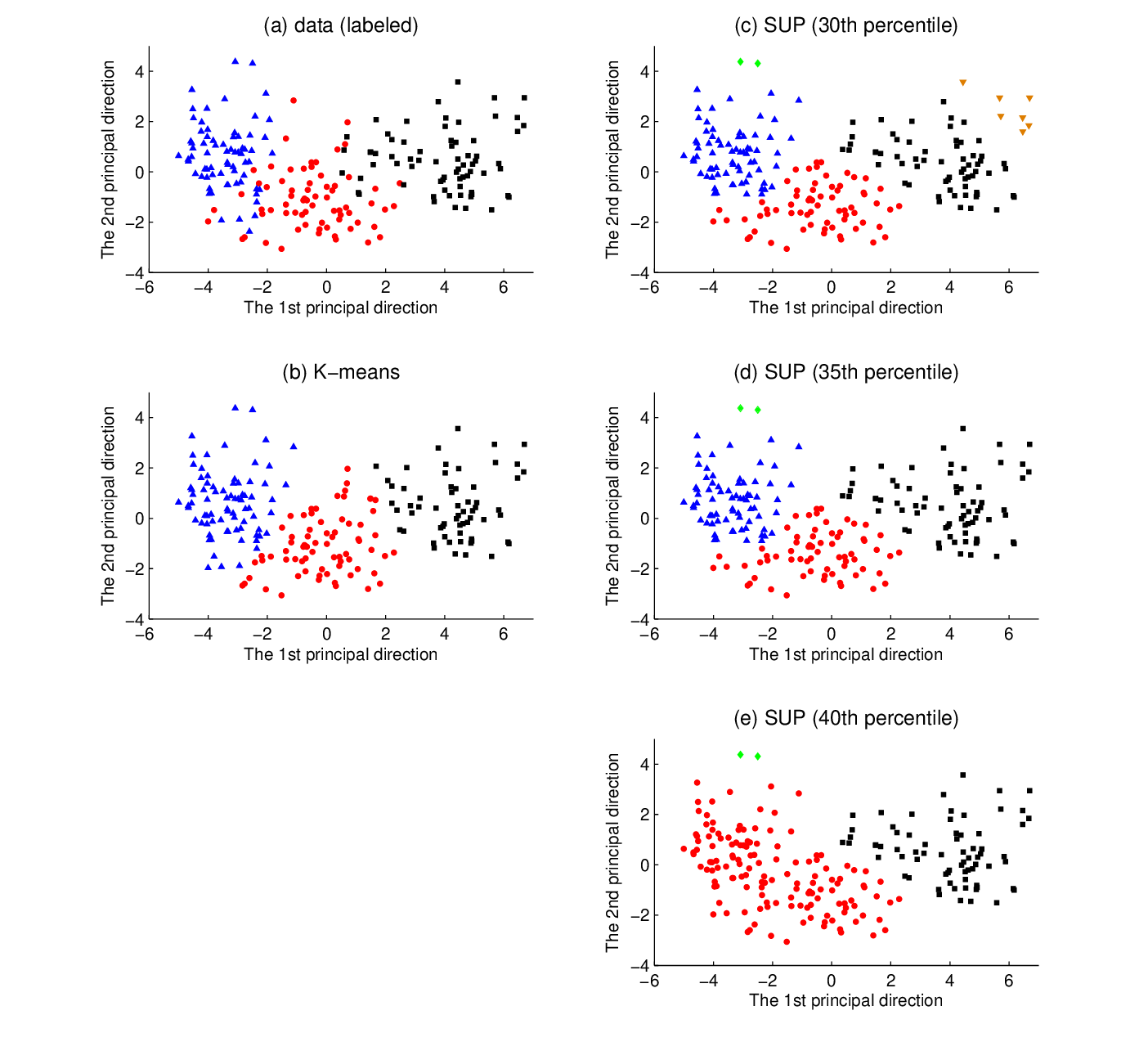}
\caption{(a) Two-dimensional projection of the seeds data. Symbols and colors
are used to distinguish between kernels from different varieties. (b)
Clustering result by the k-means. (c)-(e) Clustering results by
the dynamic SUP.} \label{fig:seeds_comp}
\end{figure}

The seeds data set is from the UCI machine learning
repository \citep{UCI}. This data has information on the geometrical
properties of kernels belonging to three different varieties of
wheat. There are $70$ kernels from each of the three varieties, and the
geometrical properties of each kernel were measured by seven
real-valued continuous variables. After principal component
analysis, the first two components explained 99.3$\%$ of the total
variation. It is therefore possible to visualize the clustering results
by a two-dimensional presentation.
Figure \ref{fig:seeds_comp}(a) shows the two-dimensional projection of
the seeds data along the first two principal directions.
Note that this data shows no clear boundary between varieties.

Figures \ref{fig:seeds_comp}(c)-(e) present the clustering results by the
dynamic SUP. The pairwise distances between kernels were used to select the $r$ values.
The frequency polygon showed a valley at around the $30$th percentile of the pairwise distance,
which produced five clusters. When the $r$ value was increased to
the $35$th and the $40$th percentiles, the dynamic SUP produced
three and two clusters, respectively, without counting the tiny
green cluster of size two that can be considered outliers.

We use the clustering results by the k-means algorithm as a comparison to present the
effect of the dynamic SUP on the kernels located at the boundaries of the varieties.
Figure \ref{fig:seeds_comp}(b) shows the
clustering results by the k-means algorithm. The major difference
between Figure \ref{fig:seeds_comp}(b) and
\ref{fig:seeds_comp}(d) is the eight points near coordinates
$(1,1)$ and the four points near coordinates $(-4,-2)$. Although we
see that the gap between the black and red clusters in Figure
\ref{fig:seeds_comp}(d) is slightly more obvious than that in Figure
\ref{fig:seeds_comp}(b), we leave it to readers' judgement.

Merging the tiny green cluster into the navy blue, we used the results presented in Figure
\ref{fig:seeds_comp}(d) that shows a structure of three clusters to calculate the validation criteria.
Two criteria were considered: the sum of the within-cluster variations and the mis-clustering errors compared
to the variety information. The sum of the within-cluster variation is $632.60$. This value is larger than $587.32$ by the k-means algorithm as expected,
because the k-means was developed to achieve the minimum value of this sum. The number of the mis-clustering errors is 22 (9.52$\%$).
The k-means produced two more errors than the dynamic SUP.

\subsection{Golub Data}
\label{subsec: golubdata} We use the gene expression data presented
in Golub {\it et al.} \cite{Golub} to show the strength of the self-updating process that can isolate noises.
The data we obtained was from the package ``multtest" (version 2.8.0) in Bioconductor. This
data contains pre-processed and normalized expression values of
$3105$ genes from $38$ patients, among whom $27$ were with acute
lymphoblastic leukemia (ALL) and $11$ were with acute myeloid
leukemia (AML). The following clustering analyses include (i)
discover gene patterns that are mostly associated with ALL-AML
distinction, and (ii) classify the $38$ patients using the $50$ genes selected by Golub {\it et al.} \cite{Golub}.

\subsubsection{Discover gene patterns}

Before we performed the self-updating process on genes,
the expression values were first normalized by genes to ensure an equal weight of each gene.

\begin{figure}[b]
\centering
\includegraphics[width=\textwidth]{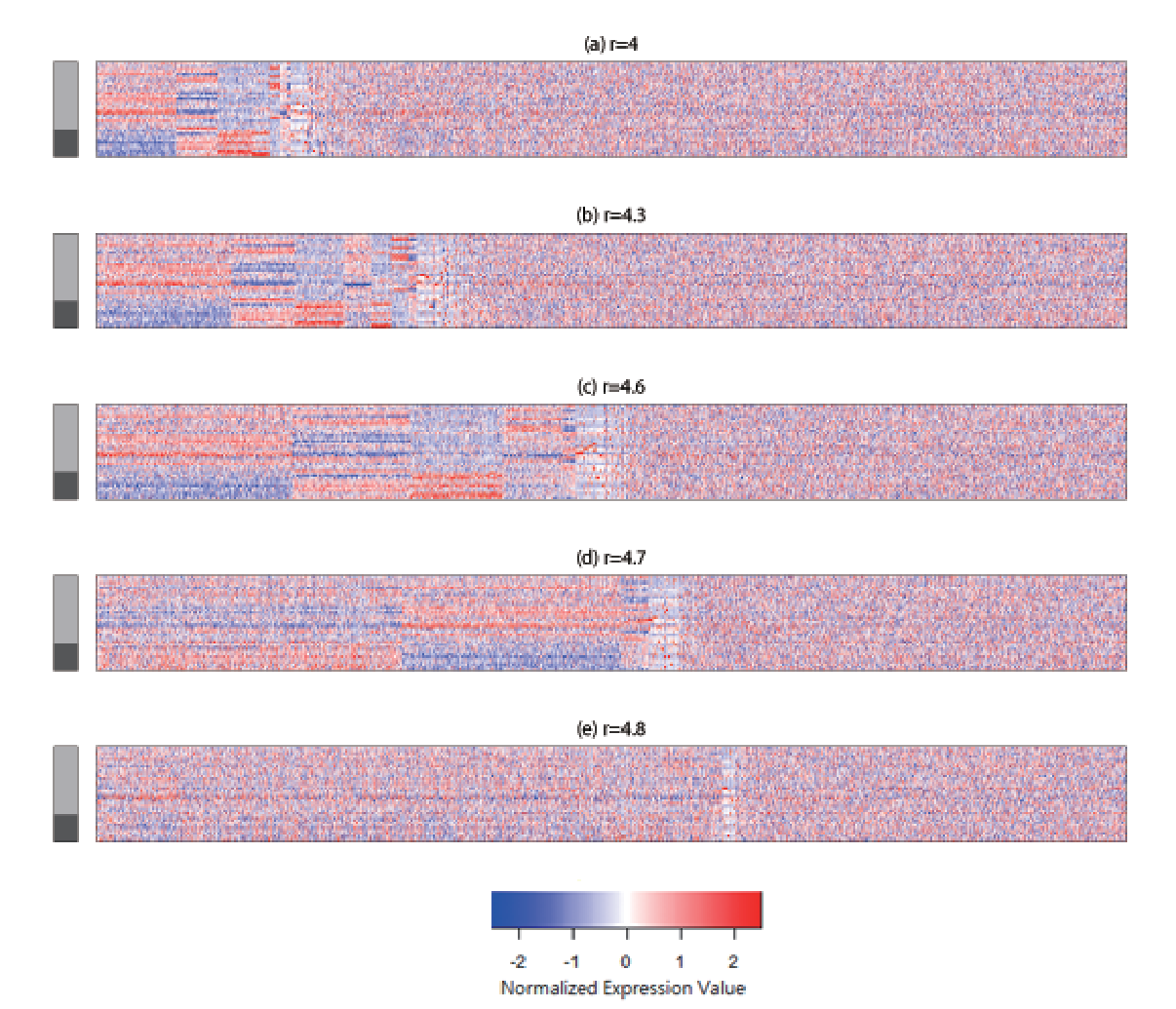}
\caption{The clustering results by dynamic SUP with different values
of $r$. The side-bar next to each of the heat-map indicates
patients' cancer types, where light and dark gray colors represent
ALL and AML patients, respectively. The heatmaps display normalized
gene expression values of $3051$ genes from $38$ patients, with each
row representing a patient's gene expression profile.}
\label{fig:heatmap_sup}
\end{figure}

The frequency polygon was not useful for the selection of the influential range $r$,
because the empirical distribution of the pairwise distances calculated from the normalized expression values did not have any sharp valley.
An alternative was to use different percentiles of the pairwise distances.
Figure \ref{fig:heatmap_sup} presents five heatmaps, which show the clustering results by the dynamic SUP using different $r$ values.
In each heatmap, gene clusters are presented from left to right according to the sizes of the clusters.
Note a large number of tiny clusters in the heatmaps.
They represent genes that showed no resemblance to other genes therefore were isolated as single-gene clusters.
These genes are often called scattered noise genes.
Figure \ref{fig:heatmap_sup} also shows that a larger r value produced fewer gene clusters,
which generally had larger sizes and larger within-cluster heterogeneity.
When $r=4.8$, SUP produced only one very large cluster; the rest were tiny clusters of scattered noise genes.

We summarize the clustering results by taking $r=4.6$ as an example.
The use of $r=4.6$ produced a total of $1478$ clusters. Among these clusters, only $9$ clusters
contained more than ten genes, and there were as many as $1420$ single-gene clusters.
In Figure \ref{fig:heatmap_sup}(c), the four largest clusters are presented from left to right, containing $580$, $349$, $276$, and $176$ genes,
respectively. These clusters showed clear gene patterns,
because SUP separated the scatter noise genes from them.
The largest cluster showed a pattern that corresponded to genes having
expression values above the mean (colored in red) for most of the
ALL patients and below the mean (colored in blue) for most of the
AML patients. The third largest cluster, in contrast, showed a pattern that corresponded to genes having high
expression values (colored in red) for most of the AML patients and low values (colored in blue) for most of the ALL patients.
To validate these clustering results,
we examined the $50$ genes that were identified to be most highly correlated with ALL-AML distinction \cite{Golub}.
Among the $50$ genes, $25$ of them were in the largest cluster, $24$ in the third largest
cluster and $1$ in the second largest cluster. When $r=4.7$ was
used, we located all of the $50$ genes in the two largest clusters,
with $25$ genes in each of the two.

\begin{figure}[htb]
\centering
\includegraphics[width=\textwidth]{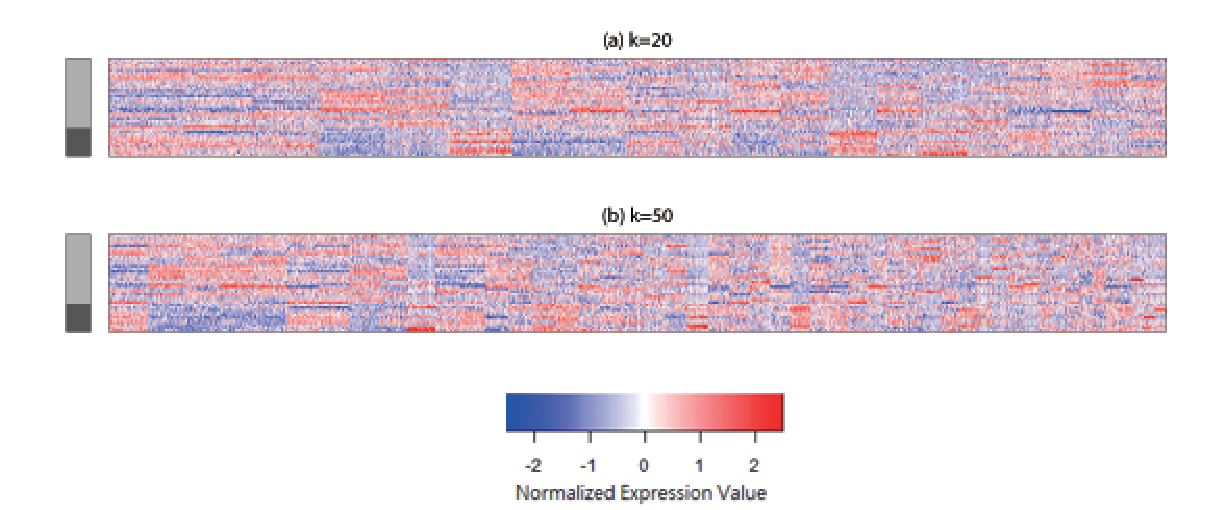}
\caption{Clustering results by the k-means with different values of
$k$'s.} \label{fig:heatmap_kmeans}
\end{figure}

In contrast to the isolations of the scattered noise genes,
Figure \ref{fig:heatmap_kmeans} presents clustering results by the k-means
using $k=20$ and $k=50$, showing that the clusters produced by the k-means were
of similar sizes. The aforementioned $50$ significant genes
were scattered to nine and twelve clusters when $k=20$
and $k=50$, respectively. These results show that meaningful gene
expression patterns were difficult to be discovered by the k-means
algorithm when the scattered noise genes are present.
The inability to handle noise is a well-known weakness of most of
the clustering methods.

\subsubsection{Classify patients}
We used the expression values of the aforementioned $50$ genes to perform
the self-updating process on the $38$ patients.
The purpose of this analysis is only to illustrate the clustering performance.

\begin{figure}[htb]
\centering
\includegraphics[width=0.5\textwidth, height=1.5in]{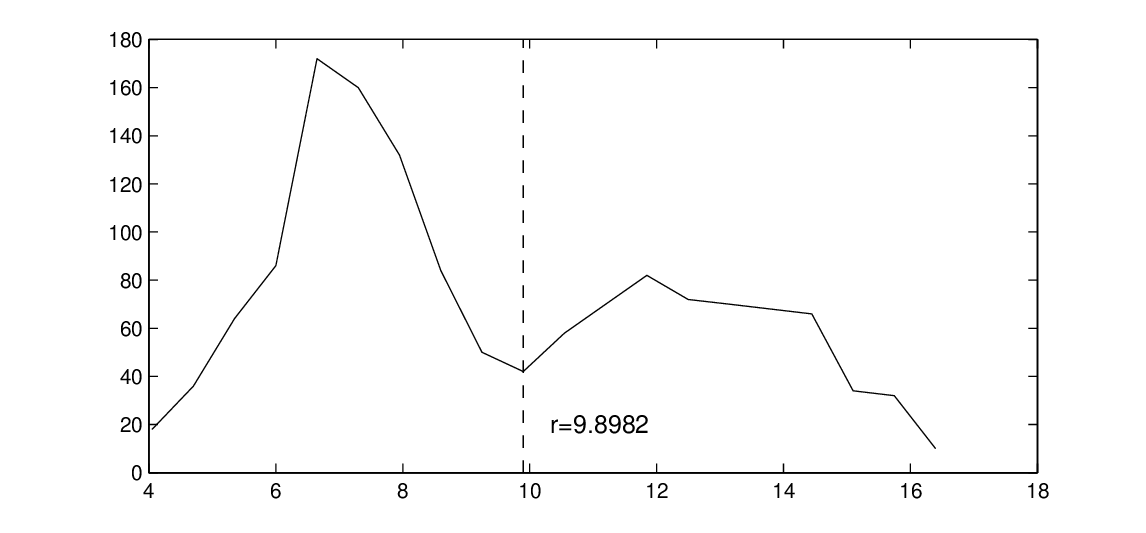}
\caption{The frequency polygon of pairwise distances between
patients} \label{fig:50_hist}
\end{figure}

Figure \ref{fig:50_hist} shows the frequency polygon of the pairwise distances between patients,
in which a sharp valley is at around $9.9$. Using this $r$ value, the dynamic SUP
produced two clusters of sizes $27$ and $11$.
These two clusters identically corresponded to the leukemia patients of two types,
meaning that SUP made a perfect distinction between the ALL and AML patients.

\section{Discussion and Conclusion}
\label{sec: discussion}
The self-updating process (SUP) is a simple,
intuitive and powerful clustering algorithm. In the updating
process, each data point moves to a new position at each iteration.
The new position depends on the influences the data point receives
from other points. At the end of the process, every point reaches
an equilibrium position without further movements. Data points that
arrive at the same position are considered to belong to the same
cluster. In this paper, we outlined the similarities and the differences between the self-updating process and the blurring mean-shift algorithm,
showing that blurring mean-shift is a static SUP.
We studied the benefits of using the self-updating process in three types of data sets.
Some of the benefits therefore also apply to the use of the blurring mean-shift for clustering.
Unless otherwise noted,
in the following we use the self-updating process to refer to both the static and the dynamic SUP, therefore including the blurring mean-shift.

Although the self-updating process was not originally developed to handle noise in data,
this ability comes naturally as a byproduct. Recall that data points
move around the sample space according to the mutual influences,
which are defined to be larger when two points are closer.
Noise data points that are not close to most of the data points therefore are to be isolated at the end of the process.
This is one strength of the self-updating process that can separate out the noise data points.
The strength offers a great advantage especially when the noise level in the data is high.

The phenomenon that a movement of a data point is more affected by its neighboring points suggests that the self-updating
process utilizes more of the local information. Clustering result of a data point can hardly be
affected by clusters relatively distant apart, regardless of how many clusters are in the data.
This property of local updating makes the self-updating process an efficient method when the number of clusters in data is large.
On the other hand, clustering algorithms such as the k-means and the fuzzy c-means algorithms that require good initial values
usually can not produce satisfactory results when clustering data sets with a large number of clusters.

In this paper, the self-updating process is not introduced as a clustering method that optimizes a certain criterion function.
Although it is often appealing to have clustering results that represent a specific
optimal solution, there are times when the optimal solutions are not what we truly seek for.
For example, the k-means algorithm uses the criterion of the sum of within-cluster variations,
which is ideal when clusters are from a mixture of normals of the same shape.
However, when clusters have different sizes and shapes, such as the example presented in Section \ref{subsec: simu_crowded},
the k-means algorithm was not unable to correctly identify clusters.
This example illustrated that sometimes the use of criterion functions for clustering is inappropriate.
The self-updating process which performs clustering by learning the relationships between data points is able to cope with different kinds of data structures.
Moreover, the dynamic SUP that utilizes even more of the local information for clustering was shown to provide the best clustering
performance among the five algorithms we studied.

From one point of view, the self-updating process is a ``slow" version of the
agglomerative hierarchical clustering: two data points are
merged ``gradually" by the self-updating process instead of ``at once" by the hierarchical clustering.
One weakness of the hierarchical clustering is that early mistakes cannot be corrected.
Slowing down the merging process especially at the beginning stage can very often
reduce the chances of making mistakes. We later realized that the
connection between the agglomerative hierarchical clustering and the self-updating process
is even closer: Consider the
following influence function:
\begin{equation}\label{eq:hm}
f_t(x_i^{(t)},x_j^{(t)})= \left \{
\begin{array}{cl}
1, & \quad d(x_i^{(t)},x_j^{(t)}) \leq r^{(t)},  \\
0, & \quad \mbox{otherwise,}
\end{array}
\right.
\end{equation}
where the influential range $r^{(t)}$ changes at each iteration,
\[
r^{(t)}=\displaystyle{\min_{ x_k^{(t)} \ne x_l^{(t)}}} ||x_k^{(t)}-x_l^{(t)}||. \\
\]
This function $f_t$ (\ref{eq:hm}) takes a positive value only when (i) $x_i^{(t)} =
x_j^{(t)}$ or when (ii) the distance
between $x_i^{(t)}$ and $x_j^{(t)}$ is the smallest among all
non-zero pairwise distances. Using this $f_t$, the self-updating process at the
first iteration only updates the pair that has the smallest pairwise
distance, and both data points of the pair are updated to the averaged
position of the pair according to (\ref{eq:update}). At later
iterations, the self-updating process only updates the two groups that have the
smallest between-group distance, and each data point in the two groups
is updated to the averaged position of all points in the two groups.
These descriptions explain that when the $f_t$ function ({\ref{eq:hm}) is considered,
the self-updating process ia identical to the agglomerative hierarchical clustering with centroid linkage.

With a proper choice of the influence function $f_t$, the self-updating process can be taken as a clustering
method that minimizes a criterion function.
For example, the $\gamma$-SUP that uses the $q$-Gaussian as the weight function is
a clustering method that minimizes the $\gamma$-divergence to the empirical data \citep{gamma_sup}.
When there is information on data structure, we can incorporate the information in the function $f_t$;
then the self-updating process can be taken as a model-based clustering method.
This is to say that with different formations of $f_t$'s, the self-updating process can have diverse looks.
Thus, learning to design $f_t$ accordingly to uncover clusters with certain structures will be our next task.

\bibliographystyle{gSCS}
\bibliography{sup}

\end{document}